\documentclass[aps,prd,twocolumn,showpacs,preprintnumbers,superscriptaddress,amsfont,graphicx,nofootinbib]{revtex4-1}
\usepackage{graphicx,amsmath,amssymb}
\usepackage{color,epsfig}

\def\bm#1{\mbox{\boldmath$#1$\unboldmath}}
\newcommand{\ie}{{\it i.e.}}

\newcommand{\M}{\mathcal{M}}
\renewcommand{\P}{\mathcal{P}}
\newcommand{\V}{\mathcal{V}}

\newcommand{\g}{\gamma}
\newcommand{\s}{\sigma}
\newcommand{\e}{\varepsilon}

\newcommand{\bra}[1]{\langle#1|}
\newcommand{\ket}[1]{|#1\rangle}

\newcommand{\SM}{\text{SM}}
\newcommand{\ct}{c_W}
\newcommand{\st}{s_W}

\newcommand{\be} {\begin{equation}}
\newcommand{\ee} {\end{equation}}
\newcommand{\ba} {\begin{eqnarray}}
\newcommand{\ea} {\end{eqnarray}}
\newcommand{\no} {\nonumber}

\begin{document}

\preprint{CERN-PH-TH-2015-160}
\preprint{FTUAM-15-19}
\preprint{IFT-UAM/CSIC-15-068} 
\preprint{ZU-TH 23/15}

\title{Probing the Charm Yukawa Coupling in Higgs + Charm Production}

\author{Ilaria Brivio} 
\email{ilaria.brivio@uam.es}
\affiliation{ 
Departamento de F\'isica Te\'orica and Instituto de F\'{\i}sica Te\'orica, IFT-UAM/CSIC,\\
Universidad Aut\'onoma de Madrid, Cantoblanco, 28049, Madrid, Spain}
\author{Florian Goertz}
\email{florian.goertz@cern.ch}
\affiliation{Theory Division, 
CERN, 1211 Geneva 23, Switzerland}
\author{Gino Isidori}
\email{isidori@physik.uzh.ch}
\affiliation{Physik-Institut, Universit\"at Z\"urich, CH-8057 Z\"urich, Switzerland}
\affiliation{INFN, Laboratori Nazionali di Frascati, I-00044 Frascati, Italy}

\date{\today}


\begin{abstract}
We propose a new method to determine the coupling of the Higgs boson to charm quarks, 
via Higgs production in association with a charm-tagged jet: $pp\to hc$.  As a first estimate, we find that 
at the LHC with 3000\,fb$^{-1}$
it should be possible to derive a constraint of order one, relative to the SM value
of the charm Yukawa coupling. As a byproduct of this analysis, we present an  
estimate of the exclusive $ pp \to hD^{(*)}$ electroweak cross section. Within the SM, the latter 
turns out to be not accessible at the LHC even in the high-luminosity phase.
\end{abstract}

\maketitle

\section{Introduction}
While the Yukawa couplings of the heavy third-generation fermions to the Higgs boson
can be measured at the LHC with a ${\cal O}$(10\,\%) accuracy, see e.g. Ref.~\cite{Peskin:2013xra}, 
constraining the diagonal Yukawa couplings of the second (first) generation quarks at a level close to the 
Standard Model (SM) expectation is very challenging. An interesting possibility, especially for the second generation, 
is trying to indirectly access these couplings via the 
radiative decays $h \to   \M + \gamma (Z)$~\cite{Bodwin:2013gca,Kagan:2014ila,Modak:2014ywa,Koenig:2015pha},
where $\M$ is a quarkonium state.\footnote{For 
indirect bounds on first generation Yukawa couplings see Ref.~\cite{Goertz:2014qia,Altmannshofer:2015qra}.}
As pointed out in Ref.~\cite{Isidori:2013cla}, the exclusive $h \to   \M V$ decays ($V=\gamma,Z,W$) 
may indeed be accessible at the SM level at the LHC and represent a precious source of information on physics beyond the SM.  
In the specific case of the charm Yukawa coupling ($Y_c$),
it should be possible to obtain bounds 2--3 times larger than the SM value in the high-luminosity (HL) phase of the LHC~\cite{Perez:2015lra}. These constraints are driven mainly by the direct search for $h\to c\bar c$ and, to a smaller extent, also by the indirect sensitivity via $h \to  J/\Psi \gamma $.

In this paper, we propose a new method to measure $Y_c$ by means of Higgs production in association with a 
 charm-tagged jet.
A particular advantage of this method, compared to the search for $h \to c \bar c$, 
lies in the fact that we probe $Y_c$ in  production --via the interaction with a charm quark from the abundant $gc$ initial state-- 
allowing to reconstruct the Higgs from its clean decay modes ($h\to \gamma\gamma$ or $h\to WW$).
This procedure strongly reduces the problem of the non-Higgs background, compared to   $h\to c\bar c$.
Moreover, requiring a single $c$-tagged jet in the final state allows to adopt high-purity (and low-efficiency) 
$c$-tag algorithms in order to reduce background (mainly from $b$-quark jets), compared to the 
case of two $c$-tagged jets (as in $h\to c\bar c$).  

Compared to the indirect sensitivity to $Y_c$  in $h \to J/\Psi \gamma$, our new method has the 
advantage of being sensitive to $Y_c$ at the tree-level and being based on a process that, 
after charm- and Higgs-tagging efficiencies, yields $O(1000)$ signal events at the HL-LHC. 
For comparison, we recall that 
${\cal B}(h \to J/\Psi  \gamma \to \mu^+\mu^- \gamma)\sim 10^{-7}$,
corresponding to $O(10)$ signal events in $pp$ collisions at 14~TeV with 3000\,fb$^{-1}$.
The main limiting factor of our approach is the theoretical uncertainty on 
$\sigma(pp\to hc)$, as a function of $Y_c$. This error could be reduced in the future by means
of higher-order QCD calculations of the ratio  $\sigma(pp\to hc)/\sigma(pp\to hb)$ as a function of 
$Y_c$ and $Y_b$. 

In principle, the production of the Higgs boson in association with a charm jet (or a charm hadron) can also 
proceed via electroweak interactions, with the charm being produced by a  real or virtual $W$ boson.
To complement this analysis, and previous studies of exclusive hadronic Higgs decays~\cite{Isidori:2013cla,Bodwin:2013gca,Kagan:2014ila,Koenig:2015pha},
we present here the first estimate of the electroweak production of the Higgs boson in association 
with a single $D$ or $D^*$  meson ($q\bar q   \to h D^{(*)}$). These processes are insensitive to the charm Yukawa coupling 
and could have represented a potential background for  the extraction of $Y_c$.
We have analyzed them in generic extensions of the SM, along the lines of Ref.~\cite{Isidori:2013cla}. 
We find that, within the SM, the exclusive electroweak production should not be visible at the LHC, even 
in the high-luminosity phase.  Moreover, we find that these process are not competitive with the corresponding 
exclusive Higgs decays  ($h\to  \M V$) as far as generic new physics (NP) searches are concerned. 
 
 The paper is organized as follows. In Sect.~\ref{sec:setup} we  introduce the setup  to describe Higgs physics 
with modified Yukawa couplings. The QCD-Yukawa $p p \to h c$ process and the corresponding extraction of $Y_c$ is discussed 
in Sect.~\ref{sec:QCDY}. The exclusive electroweak $pp\to  h D^{(*)}$ process  is analyzed in Sect.~\ref{sec:EW}.
The results are summarized in the Conclusions.

\section{Setup}
\label{sec:setup}
Within the SM the couplings of the physical Higgs boson to the fermions are completely determined in terms of  fermion masses. 
However, in the presence of NP, a misalignment between quark-mass and Yukawa matrices is possible.
This can be parametrized in a model-independent way by adding the $D=6$ operators	
	\begin{equation}
	\label{eq:D6}
	{\cal L}_6^Y =- \frac{1}{v^2}\left( (\Phi^\dagger \Phi)\,\bar q_L \bm{C}_u \Phi^c u_R +  (\Phi^\dagger \Phi)\, \bar q_L 
	\bm{C}_d \Phi\,d_R\right)\,
	\end{equation}
to the SM Lagrangian. Here, $\Phi$ denotes the Higgs doublet, parametrized in unitary gauge as $\Phi=1/\sqrt 2 \left(0,h+v\right)^T$,
where $v$ corresponds to the vacuum expectation value $\langle \Phi \rangle = 1/\sqrt 2\,\left(0,v\right)^T $, $h$ is the physical
Higgs field, and $q_L, u_R, d_R$ are the chiral SM-quark doublet and singlets (all quark fields being 3-vectors in flavor space). 
Inserting this decomposition of the Higgs doublet into (\ref{eq:D6}) as well as into the SM-like ($D=4$) Yukawa terms 
with couplings $\hat{\bm{Y}}^{u,d}_{\rm SM}$, we obtain the fermion masses and Higgs couplings in the flavor basis
	\begin{equation}
	\label{eq:LM}
	{\cal L} \supset - \bar u_L \left( \hat{\bm{M}}^u + \frac{h}{\sqrt 2} \hat{\bm{Y}}^u \right) u_R
	- \bar d_L \left( \hat{\bm{M}}^d +  \frac{h}{\sqrt 2} \hat{\bm{Y}}^d \right)  d_R\, ,
	\end{equation}
where the Yukawa matrix $\hat{\bm{Y}}^{u,d} =\hat{\bm{Y}}^{u,d}_{\rm SM} + \frac 3 2\, \bm{C}_{u,d}$ and 
the mass matrix $\hat{\bm{M}}^{u,d} =\frac{v}{\sqrt 2}(\hat{\bm{Y}}^{u,d}_{\rm SM} + \frac 1 2\bm{C}_{u,d})= \frac{v}{\sqrt 2}
( \hat{\bm{Y}}^{u,d} -  \bm{C}_{u,d})$ are independent parameters. 
After performing a rotation to the mass basis
	\begin{equation}
	\begin{split}
	\hat{\bm{M}}^u &= \bm{U}_L^u\, \bm{M}_{\rm diag}^u \bm{U}_R^{u\, \dagger},
	\ \, \bm{M}_{\rm diag}^u\!={\rm diag}(m_u,m_c,m_t)\,,\\
	\hat{\bm{M}}^d &= \bm{U}_L^d\, \bm{M}_{\rm diag}^d \bm{U}_R^{d\, \dagger},
	\ \, \bm{M}_{\rm diag}^d\!={\rm diag}(m_d,m_s,m_b)\,,
	\end{split}
	\end{equation}
with $\bm{U}_L^d = \bm{U}_L^u\, \bm{V}_{\rm CKM}$,
we finally arrive at the couplings of the physical quarks to the Higgs boson
 	${\bm{Y}}^u = \bm{U}_L^{u\, \dagger}  \hat{\bm{Y}}^u \bm{U}_R^u$,
	${\bm{Y}}^d =  \bm{U}_L^{d\, \dagger} \hat{\bm{Y}}^d  \bm{U}_R^d$,
such that
	\begin{equation}
	{\cal L} \supset - \bar u_L \left( \bm{M}_{\rm diag}^u+ \frac{h}{\sqrt 2} \bm{Y}^u  \right) u_R\, 
	+\, (u \to d).\\
	\end{equation}

Here, we concentrate on possible experimental constraints on the diagonal entry $Y_c \equiv (\bm{Y}^u)_{22}$.
 For convenience, we parametrize the deviations from the SM prediction ($\bm{C}_u=\bm{C}_d=0$) in terms of 
$\kappa_q \equiv Y_q v/(\sqrt 2 m_q) \neq 1$, which we assume to be real for simplicity.\footnote{
In the following we assume the top and bottom Yukawa couplings to be constrained close to their SM values
after the high-luminosity LHC run.} 

\section{The QCD-Yukawa $p p \to h c$ process}
\label{sec:QCDY}
\begin{figure}[!t]
	\begin{center}
	\includegraphics[height=0.935in]{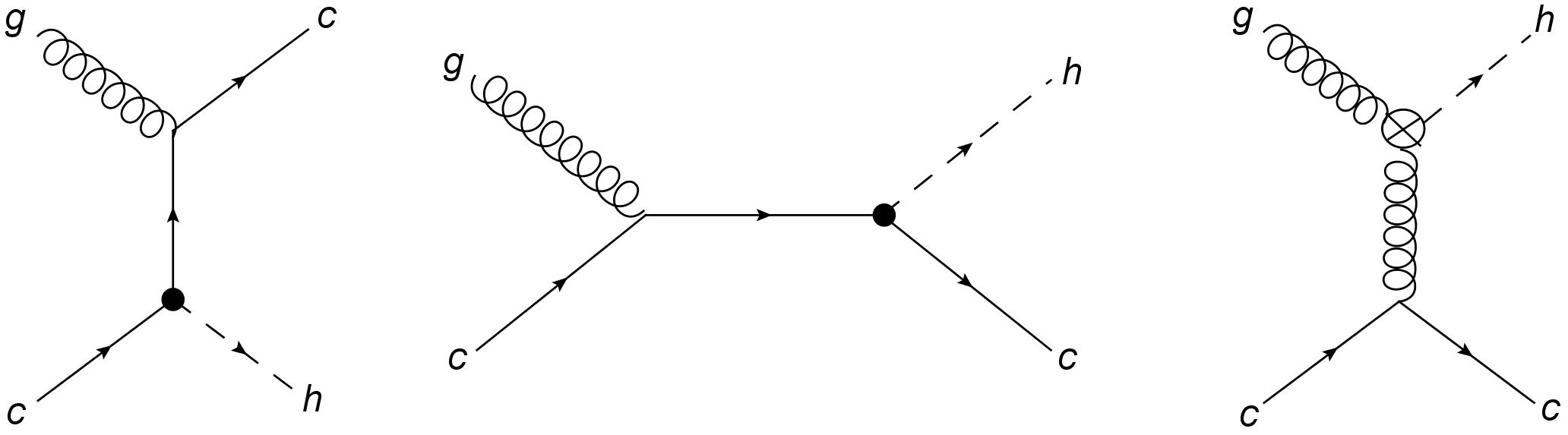}
	\caption{\label{fig:diag} Diagrams contributing to $p p \to h c$ at  leading order. Black dots correspond to vertices 
	where the Yukawa coupling $Y_c$ enters, while the crossed vertex corresponds to the SM-like top triangle, integrated out.}
	\end{center}
	\end{figure}

We consider the production of a Higgs boson in association with a charm-quark jet. At the LHC, the main partonic
process inducing this final state is $g c \to h c$ and the corresponding Feynman diagrams are presented in Figure~\ref{fig:diag}.
The charm Yukawa coupling, depicted as a black dot, enters in the first two graphs, that yield a 
contribution to the amplitude of $O(g_s Y_c)$. The $t-$channel diagram turns out to be largely dominant.
The third diagram is formally of higher order in $\alpha_s$ but is enhanced by the top-quark Yukawa coupling. Here the crossed vertex
corresponds to the effective $ggh$ interaction obtained by integrating out the top quark. This diagram yields
the contribution to the amplitude that survives in the limit $\kappa_c\to 0$ (see Table~\ref{tab:kc}).

The challenge of the proposed process is to tag the charm-quark jet, as in $h\to c\bar c$.
However, as anticipated, it offers some interesting virtues compared to $h\to c\bar c$. 
In particular, it allows us to fully reconstruct the Higgs boson in a clean decay channel such 
as $h \to \gamma \gamma$ or $h \to WW$, and it requires only a single charm tag. 
The main drawback is that the process does not vanish in the limit $Y_c \to 0$ (contrary to 
$h\to c\bar c$) requiring a good theoretical control on the cross section as a function of $Y_c$.
While a full analysis, including the optimization of the event selection, is beyond the scope of this article,
here we just want to examine the potential of the channel by deriving the expected number of 
signal and background events, based on reasonable efficiency assumptions.
 
We have calculated the cross section of $p p \to h c$ at leading order in QCD (including the effective $ggh$ as discussed above)
at the LHC with 14\,TeV center-of-mass energy for
various values of $\kappa_c$, employing {\tt MadGraph5} \cite{Alwall:2011uj}, with a tailored model file and CTEQ6L1 
parton distribution functions. Using $m_c(m_Z)=0.63$\,GeV and $m_h=125\,$GeV,
for $\kappa_c=1$ (\ie, the SM) we obtain a cross section
of $\sigma(p p \to h c)=166.1\,$fb, employing the default cuts of $p_T(j)\!>\!20\,{\rm GeV},\, 
\eta(j)\!<\!5,\, \Delta R(j_1,j_2)\!>\!0.4$ for all processes considered here.
In the following, we focus on the $h \to \gamma \gamma$ decay channel, 
with a branching fraction of ${\cal B}(h \to \gamma \gamma)= 0.0023$.
This leads to $S_0 =2292$ events at the HL-LHC with $3000\,$fb$^{-1}$, 
taking into account also the $p p \to h \bar c$ process. Assuming a charm-tagging efficiency of 
$\epsilon_c=0.4$ (see e.g.~Ref.~\cite{Perez:2015lra}), we finally end up with 
$S=\epsilon_c S_0=917$ signal events. The different number of events obtained by varying 
$\kappa_c$ are reported in Table \ref{tab:kc}.

\begin{table}
\raggedright
\hspace{8mm}
\begin{tabular}{|c||c|c|c|c|c|c|c|c|c|}
\hline
$\kappa_c$ & 0 & 0.25 & 0.5 & 0.75 & 1 & 1.25 & 1.5 & 1.75 & 2  \\
\hline
$S$ & 874 & 877 & 885 & 899 & 917 & 941 & 973 & 1008 & 1052 \\
\hline
\end{tabular}

\vspace{1.4mm}
\hspace{8mm}
\begin{tabular}{|c||c|c|c|c|c|c|c|c|c|c|}
\hline
$\kappa_c$  & 2.25 & 2.5 & 2.75 & 3 & 3.25 & 3.5 & 3.75 & 4 & 4.25 & 4.5\\
\hline
$S$ & 1097 & 1148 & 1206 & 1276 & 1350 & 1424 & 1504 & 1590 & 1683 & 1786 \\
\hline
\end{tabular}
\caption{\label{tab:kc} Number of Signal events $S(\kappa_c)$ in dependence on the charm-quark Yukawa coupling. 
See text for details.}
\end{table}

The main backgrounds to the process studied here are $p p \to h g$, 
with the gluon mis-identified as a charm quark, as well as $p p \to h b$,
with the bottom quark being mis-tagged. In the first case, we treat 
separately the case $p p \to h c \bar c$, where only one charm-quark jet is reconstructed
and the case where the gluon produces a light quark jet.
The backgrounds feature $\sigma(p p \to h g)=12.25\,$pb, $\sigma(p p \to h b)=203\,$fb,
as well as $\sigma(p p \to h c \bar c)=55\,$fb.
We employ a conservative assumptions for the jet reconstruction efficiency of 
$1-\epsilon^{\rm miss}=95\%$, 
as well as $g\to c$ and $b\to c$ mis-tag rates of $\epsilon_{\rm g \to c}=1\%$ and $\epsilon_{\rm b \to c}=30\%$.
With these figures we obtain $B=1705$ background events at $3000\,$fb$^{-1}$, leading to
$N(\kappa_c=1)=S(\kappa_c=1)+B=2622$ total events.  
We then assume a statistical error on the total number of events  $(\sqrt N)$ and a theoretical (relative) error on the 
signal events of $20\%$. The latter is deduced by the recent next-to-leading order (NLO) 
analysis of the Higgs production in association with bottom quarks~\cite{Wiesemann:2014ioa}.
Finally, statistical and theoretical error are added in quadrature.\footnote{
The two dominant backgrounds, $p p \to h b$ and $p p \to h g$, can both be directly 
measured at the LHC with specific tags (inverted $b$ vs.~$c$ tag for the former and 
light-quark-jet tag for the latter) - this is why we do not assign an additional theory error to them.
}

\begin{figure}[!t]
	\begin{center}
	\includegraphics[height=1.58in]{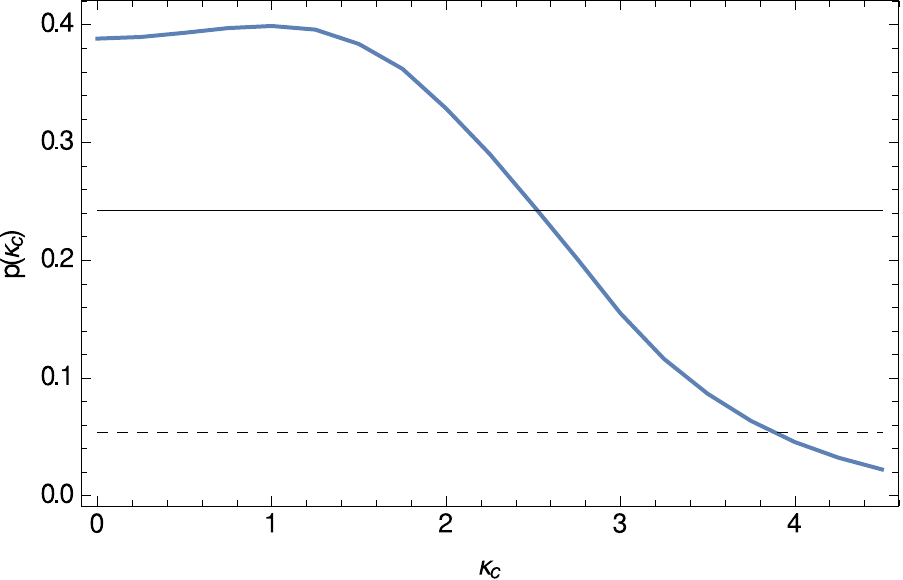}
	\caption{\label{fig:p} The expected $p$-value for a given value of $\kappa_c$ from the process $p p \to h c$ at the 14 TeV LHC with 3000\,fb$^{-1}$ and a conservative assumption for the theoretical uncertainty. 
	See text for details.}
	\end{center}
	\end{figure}

In the following, we want to examine the expected constraints that can be set on $\kappa_c$ from the process under
consideration. To this purpose, we assume the SM to be true and calculate how many standard deviations $\Delta N(\kappa_c)$ 
away a prediction $N(\kappa_c)$ is from $N(\kappa_c=1)$, which is the expected outcome of the experiment. 
The values of $\kappa_c$ that lead to a discrepancy of more than $n$ standard deviations are then expected to be 
excluded at $n\, \sigma$. We plot the corresponding $p$-value, $p(\kappa_c)$, in Figure~\ref{fig:p}
approximating the Poisson distribution of the number of events by a Gaussian. 
The $1\sigma$ and $2\sigma$ equivalents are depicted by the solid and dashed lines, respectively. A conservative estimate for the expected 1-$\sigma$ ($95\%\,$CL) constraint on $\kappa_c$ is thus obtained as 
\begin{equation}
|\kappa_c| < 2.5\ (3.9),
\end{equation}
which lies in the ballpark of the results quoted in \cite{Perez:2015lra},
where the latter combines ATLAS and CMS to arrive at $2 \times 3000\,$fb$^{-1}$ of integrated luminosity.

On the other hand, an improved prediction of the SM cross section $\sigma(p p \to h c)$, leading to $\delta_{\rm th}=10\%$, would strengthen our expected 1-$\sigma$ ($95\%\,$CL) limit to
\begin{equation}
|\kappa_c| < 1.9\ (2.6),
\end{equation}
approaching the SM value of $Y_c$. 

We note that optimized cuts can still increase $S/B$ and in particular lead to an enhanced sensitivity on $\kappa_c$. 
As the statistics at 3000 fb$^{-1}$ is large enough, there are good prospects to still improve the bounds.
A corresponding detailed investigation, including detector simulation, is beyond the purpose of this letter and 
can be performed best by the experimental community.

We further stress that the dominant source of uncertainty, at present,  is the theoretical 
error on $\sigma(p p \to h c)$. We have indeed checked that  the result does not change significantly 
worsening the $g\to c$ and $b\to c$ mis-tag rates to $5\%$ and $40\%$, respectively. 
As far as the reliability (and possible reduction) on the theoretical error is concerned, 
a promising possibility would be a dedicated calculation of $\sigma(p p \to h c)/\sigma(p p \to h b)$ at NLO (or NNLO),
as a function of $Y_c/Y_b$,  supplemented by measurements of this ratio and 
$\sigma(p p \to h b)$ with a combination of normal and inverted $b$ vs.~$c$ tags.

\section{The electroweak $pp\to h\M$ process}
\label{sec:EW}

As anticipated in the introduction, the production of the Higgs boson in association with charm  
can proceed also via electroweak interactions, starting form an initial charm-less $q\bar q^\prime$ state
($u\bar d \to  h W^{(*)} \to h c\bar s$).
The case of an on-shell $W$ producing a charm jet  can  be discriminated from 
the QCD-Yukawa process   by means of appropriate cuts on the  jet momentum. 
Less obvious is the discrimination in the case of a virtual $W^*$ producing a low-momentum 
$c$-jet, or even a single charmed hadron. In the following we estimate in detail the specific case 
of the single meson production: $pp   \to h \M$,  with $\M$ being a charmed meson
or a charmonium state.

\begin{figure}[!t]
	\begin{center}
	\includegraphics[height=0.935in]{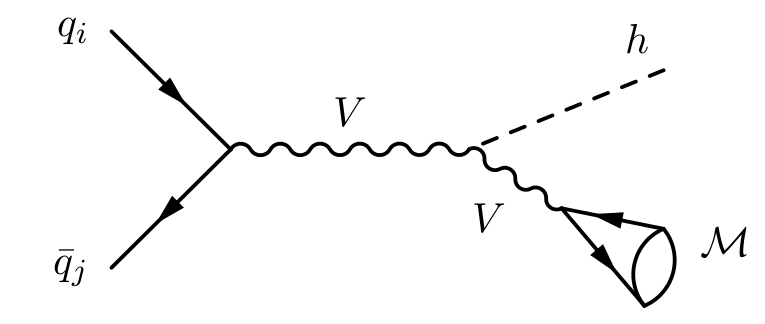}
	\caption{\label{fig:diag2} Diagram contributing to $p p \to h \M$ at  leading order, where
	 $V=W^\pm,Z$.}
	\end{center}
\end{figure}	

The leading partonic amplitude within the SM is shown in Fig.~\ref{fig:diag2}.
Following Refs.~\cite{Isidori:2013cla,Isidori:2013cga}, we  parameterize 
the quark currents appearing in the initial and final state with arbitrary 
vector and axial couplings:
\begin{equation}\label{J_Q}
  J_{q,ij}^\mu = \bar{q}^i (g_{V,ij}\, \g^\mu + g_{A,ij}\, \g^\mu \g_5)q^j\,.
\end{equation}
The matrix element of the current that generates the meson in the final state assumes one of the following structures, depending on the spin of $\M$:
\begin{equation}\label{J_M}
 \bra{\M(p,\epsilon)}J_q^\nu\ket{0}=\frac{1}{2}
 \begin{cases}
  g_\P\,f_\P\, p^\nu& 		\M\equiv\P \\[2mm]
  g_\V\,f_\V\, m_\V \,\e^\nu& 	\M\equiv\V \,,\\
 \end{cases}
\end{equation}
where $f_\M$ is the meson's decay constant,  and $g_\M$ encodes the dependence 
on the coupling to the relevant gauge boson
($g_\P= g_{A,ij}$, $g_\V=g_{V,ij}$ for a $\langle \bar q^j  q^i \rangle$ meson).\footnote{
$g_{V,uu} =   \frac{g}{\ct}\left(\frac{1}{4}-\frac{2}{3}\st^2\right)$, 
$g_{V,dd} =  \frac{g}{\ct}\left(-\frac{1}{4}+\frac{1}{3}\st^2\right)$,
$g_{A,uu} = - g_{A,dd} = - \frac{g}{4\ct}$,
$g_{V,u^i d^j}=-g_{A,u^i d^j}=  \frac{g}{2\sqrt2}\,(V_{\rm CKM})_{ij}$}
With this notation, the SM expression for the partonic cross section for the case of a pseudoscalar meson reads
\begin{equation}
\label{dS.SM.P}
 \s(q\bar{q}^\prime\to h \P)_\SM(q^2) = \frac{g_\P^2(g_{V}^2+g_{A}^2) f_\P^2 q^2  }{576\pi v^2 (q^2-m_V^2)^2}\lambda^3(q^2)\, ,
 \end{equation}
where $V=W^\pm,Z$, and we have suppressed the indices of $g_{A,V}$ for simplicity.
The vector case has the same functional form with $\P \to \V$, 
up to tiny  $\mathcal{O}\left(m_\V^2/m_V^2\right)$ corrections.
In the above expression, $q^2$ denotes the total momentum of the initial state in the partonic process and
\ba
 \lambda(q^2)&=& \sqrt{1-2\,\frac{m_h^2+m_\M^2}{q^2}+\frac{(m_h^2-m_\M^2)^2}{q^4}}\, .
 \ea
Convoluting the cross sections with the appropriate PDF in the region $130\leq \sqrt{q^2} \leq 1~{\rm TeV}$, and assuming an integrated  
luminosity of $3000$~fb$^{-1}$ we obtain the expected number of events for each channel at the HL-LHC. 
The results, summarized in Table~\ref{tab.expectedSMevents}, show that these processes will not be observable at the SM level,
and certainly do not represent a  dangerous   background for the QCD-Yukawa process discussed in Sect.~\ref{sec:QCDY}.

\begin{table}[bt]
 \centering\renewcommand{\arraystretch}{1.3}
 \begin{tabular}{|c|cc|cc|}
 \hline
 Channel & $~m_\M$ & $f_\M~$   & \multicolumn{2}{c|}{Events @ HL-LHC}\\
 & \multicolumn{2}{|c|}{ (MeV)} &  Method (a)& 		Method (b)\\\hline
 $\eta_c$ &	2984&			200 &	0.10&			0.08\\
 $J/\psi$ &	3100&			410 & 	0.08&			0.07\\
  $D_s^\pm$ &	1968&			250 & 	0.48&			0.40\\
  $D_s^{*\pm}$&2112&			325 &		0.84&			0.69\\
  \hline
 \end{tabular}
\caption{Expected number of $h\mathcal{M}$  associated production events at HL-LHC (14 TeV and $3000$~fb$^{-1}$) in the energy region $130\leq \sqrt{q^2} \leq 1~{\rm TeV}$ for representative charmed-meson final states. The results reported under Method (a) are obtained rescaling bin-by-bin the cross section distribution of Drell-Yan processes provided by MadGraph~5~\cite{Alwall:2011uj}. The computation of Method (b) is performed via numerical convolution of the analytic cross section with the PDF of the MSTW~2008 libraries~\cite{Martin:2009bu}. Both account only for SM contributions.}\label{tab.expectedSMevents}
\end{table}

Given the smallness of the SM signal, it is worth to investigate if these cross sections can be significantly altered beyond the SM.
This can be done generalizing the approach of Refs.~\cite{Isidori:2013cla,Isidori:2013cga}. 
The leading (helicity-conserving) transition amplitude can be decomposed in full generality as
\be
 \mathcal{A}(q\bar{q}^\prime \to h \M)=- J_q^\mu T_{\mu\nu}  \bra{\M }J_q^\nu\ket{0}~.
\ee 
To a good accuracy  the quark current is conserved ($q_\mu J_q^\mu =0$),  and the tensor $T_{\mu\nu}$ can be decomposed 
in terms of only four Lorentz structures. Using the same notation as in Ref.~\cite{Isidori:2013cla}:
\ba
 T_{\mu\nu} &=& f_1(q^2) g_{\mu\nu} + f_2(q^2) p_\mu p_\nu + f_3(q^2) (p\cdot q\, g_{\mu\nu}-p_\mu q_\nu) \no \\
&&  + f_4(q^2)\e_{\mu\nu\rho\s}p^\rho q^\s\,,
\ea
where $q_\mu$ is the total momentum of the quark pair in the initial state, and $p_\mu$ is the  meson momentum ($p^2=m_\M^2$).
 With these notations the partonic cross section reads
\ba
&& \s(q \bar{q}^\prime \to h \P) (q^2) = \frac{g_\P^2 f_\P^2}{2304\pi}    \left(g_{V}^2+g_{A }^2\right)  \no \\
&& \qquad \times\left|f_1(q^2)+m_\P^2 f_2 (q^2) \right|^2q^2\lambda^3(q^2)\,,
\label{dS.d8.P}
\ea
where, similarly to the SM case, $\s(q \bar{q}^\prime \to h \V)$ 
has the same functional form up to tiny  $\mathcal{O}\left(m_\V^2/m_V^2\right)$ corrections.
Neglecting the latter terms, we  obtain 
\be
 \frac{\s(q\bar{q}^\prime\to h\M)_\text{BSM}}{\s(q\bar{q}^\prime\to h\M)_\SM} (q^2) = \left| \frac{ f_1(q^2)^{\phantom{\rm SM}} }{f_1(q^2)^{\rm SM}}  \right|^2\,,
\ee
where $ f^{\rm SM}_1(q^2) \propto 1/(v(q^2-m_V^2))$ and we disregard potential changes to the fermionic
currents. Deviation from the SM are thus induced by possible 
non-pole-terms (i.e.~contact terms) in the form factor $f_1(q^2)$. Within a generic effective-field theory (EFT) approach to Higgs physics
(both linear and non-linear EFT), contact terms in $f_1(q^2)$ are generated by dimension-six operators.  
However, their effect would show-up exactly in the same functional form either 
in the on-shell associated production ($pp\to Vh$) or in $h\to V \M$ decays, that share the same 
current structure~\cite{Isidori:2013cla,Isidori:2013cga}.
Since the latter processes can be measured (or at least bounded) to a better accuracy, we conclude that 
$\s(pp\to h\M)$   is not a very sensitive probe of generic extensions of the SM.

\section{Conclusions}
\label{sec:concl}
In this letter, we proposed a new strategy for the measurement of the Yukawa coupling of the charm quark:
the measurement of the production cross section of the Higgs boson in association with a charm jet. A first estimate showed that $Y_c$
could be determined at a level approaching the SM value in this channel, which offers  virtues and 
drawbacks  quite different with respect to the $h\to c\bar c$ search. 
A fully realistic analysis was beyond the scope of the present paper. 
A more realistic evaluation of the efficiencies is likely to decrease the number of signal events $S$ compared to our naive estimate;
however, as we have discussed, the sensitivity on $Y_c$ could even increase with properly designed $b$ and $c$ tag strategies
aimed to measure the background  from data and to reduce the theoretical error on the normalization of the cross-section.
This first analysis therefore calls for more detailed studies both on the theory and on the experimental side.

\paragraph*{Acknowledgements}
We are grateful to Andreas Papaefstathiou and Dieter Zeppenfeld for useful comments. 
The research of I.B.~is supported by the Spanish MINECO’s “Centro de Excelencia Severo Ochoa” Programme under grant SEV-2012-0249, and by an ESR contract of the EU network FP7 ITN INVISIBLES (Marie Curie Actions, PITN-GA-2011-289442). F.G.~acknowledges the support of a Marie Curie Intra European Fellowship within the EU FP7 (grant no. PIEF-GA-2013-628224). The research of G.I.~is supported in part by the Swiss National Science Foundation 
(SNF) under contract 200021-159720.


\begin{thebibliography}{99}


\bibitem{Peskin:2013xra}
  M.~E.~Peskin,   arXiv:1207.2516;
  arXiv:1312.4974.




\bibitem{Bodwin:2013gca}
  G.~T.~Bodwin, F.~Petriello, S.~Stoynev and M.~Velasco,
  Phys.\ Rev.\ D {\bf 88} (2013) 5,  053003
  [arXiv:1306.5770].

\bibitem{Kagan:2014ila}
  A.~L.~Kagan, G.~Perez, F.~Petriello, Y.~Soreq, S.~Stoynev and J.~Zupan,
  arXiv:1406.1722.

\bibitem{Modak:2014ywa}
  T.~Modak and R.~Srivastava,
  arXiv:1411.2210.

\bibitem{Koenig:2015pha}
  M.~Koenig and M.~Neubert,
  arXiv:1505.03870.
  
  

\bibitem{Goertz:2014qia}
  F.~Goertz,
  Phys.\ Rev.\ Lett.\  {\bf 113} (2014) 261803
  [arXiv:1406.0102].
  

\bibitem{Altmannshofer:2015qra}
  W.~Altmannshofer, J.~Brod and M.~Schmaltz,
  JHEP {\bf 1505} (2015) 125
  [arXiv:1503.04830].
  
\bibitem{Isidori:2013cla}
  G.~Isidori, A.~V.~Manohar and M.~Trott,
  Phys.\ Lett.\ B {\bf 728} (2014) 131
  [arXiv:1305.0663].
 
  
\bibitem{Perez:2015lra}
  G.~Perez, Y.~Soreq, E.~Stamou and K.~Tobioka,
  arXiv:1503.00290;
  arXiv:1505.06689.

\bibitem{Alwall:2011uj}
  J.~Alwall, M.~Herquet, F.~Maltoni, O.~Mattelaer and T.~Stelzer,
  JHEP {\bf 1106} (2011) 128
  [arXiv:1106.0522].
 
\bibitem{Wiesemann:2014ioa}
  M.~Wiesemann, R.~Frederix, S.~Frixione, V.~Hirschi, F.~Maltoni and P.~Torrielli,
  JHEP {\bf 1502} (2015) 132
  [arXiv:1409.5301 [hep-ph]].


\bibitem{Isidori:2013cga}
  G.~Isidori and M.~Trott,
  JHEP {\bf 1402} (2014) 082
  [arXiv:1307.4051].
   
  
  
\bibitem{Martin:2009bu}
  A.~D.~Martin, W.~J.~Stirling, R.~S.~Thorne and G.~Watt,
  Eur.\ Phys.\ J.\ C {\bf 64} (2009) 653
  [arXiv:0905.3531];
  Eur.\ Phys.\ J.\ C {\bf 63} (2009) 189
  [arXiv:0901.0002].
 
 
  
\end{thebibliography}
\end{document}